\documentclass[preprintnumbers,10pt,nofootinbib]{revtex4}
\pdfoutput=1

\usepackage{amsmath,latexsym,amssymb,amsfonts}
\usepackage[pdftex]{color,graphicx}
\usepackage{bm}

\begin{document}


\title{\textbf{Charged black holes in string-inspired gravity}\\
\small{II. Mass inflation and dependence on parameters and potentials}}

\author{\textsc{Jakob Hansen$^{1}$}\footnote{{\tt hansen{}@{}kisti.re.kr}} and \textsc{Dong-han Yeom$^{2}$}\footnote{{\tt innocent.yeom{}@{}gmail.com}}}

\affiliation{\textit{$^{1}$\small{KISTI, Daejeon 305-806, Republic of Korea}}\\
\textit{$^{2}$\small{Leung Center for Cosmology and Particle Astrophysics, National Taiwan University, Taipei 10617, Taiwan}}
}

\begin{abstract}
We investigate the relation between the existence of mass inflation and model parameters of string-inspired gravity models. In order to cover various models, we investigate a Brans-Dicke theory that is coupled to a $U(1)$ gauge field. By tuning a model parameter that decides the coupling between the Brans-Dicke field and the electromagnetic field, we can make both of models such that the Brans-Dicke field is biased toward strong or weak coupling directions after gravitational collapses. We observe that as long as the Brans-Dicke field is biased toward any (strong or weak) directions, there is no Cauchy horizon and no mass inflation. Therefore, we conclude that to induce a Cauchy horizon and mass inflation inside a charged black hole, either there is no bias of the Brans-Dicke field as well as no Brans-Dicke hair outside the horizon or such a biased Brans-Dicke field should be well trapped and controlled by a potential.
\end{abstract}

\maketitle

\newpage

\tableofcontents

\section{Introduction}

Motivated from quantum gravity \cite{DeWitt:1967yk,Green:1987sp,Kiefer} as well as cosmology \cite{Kachru:2003aw} and black hole physics \cite{Maldacena:1997re}, many models of modified gravity have been suggested. One big branch of the modified gravity is the scalar-tensor gravity \cite{Brans:1961sx,Fujii:2003pa} that can be motivated from string theory \cite{Becker:2007zj}. In these classes of string-inspired models, the investigation of black hole physics is very worthwhile, especially not only for static solutions \cite{Gibbons:1987ps}, but also for dynamical behaviors.

For this purpose, in the authors previous paper \cite{Hansen:2014rua}, we investigated the prototype of the string-inspired model as
\begin{eqnarray}
S = \frac{1}{16\pi} \int \sqrt{-g}d^{4}x \left[\Phi R - \frac{\omega}{\Phi}\Phi_{;\mu}\Phi_{;\nu}g^{\mu\nu} - \Phi^{\beta} F^{2} \right],
\end{eqnarray}
where $R$ is the Ricci scalar, $\Phi$ is the Brans-Dicke field that can be interpreted as a dilaton field (if $\omega = -1$, while there can be other string-inspired values for $\omega$ \cite{Garriga:1999yh}), and $F$ is the two-form field that can couple to a complex scalar field. In order to investigate dynamical properties, we implemented a numerical formalism, the double-null formalism \cite{Hamade:1995ce}, and people have obtained various results \cite{doublenull,Yeom1}. Especially, in the authors' first paper, we mainly focused to see the causal structures and responses of the Brans-Dicke field $\Phi$.

One important result of the authors' previous paper was to related causal structures of string-inspired charged black holes (see also \cite{Yeom3}) with the responses of the Brans-Dicke field \cite{Scheel:1994yr,Yeom2}. Because of a certain coupling between the gauge field and the Brans-Dicke field, the responses of the Brans-Dicke field \textit{during} and \textit{after} a gravitational collapse can be sensitively changed. Because of this, the internal causal structure of a charged black hole could be sensitively depend on the choice of coupling parameters $\omega$ and $\beta$, whether there exists a Cauchy horizon and mass inflation \cite{Poisson:1990eh} or not. In conclusion, we observed that if $\omega > -3/2$ and $\beta > 0$, then there was no Cauchy horizon, where this result is consistent with \cite{Ann}.

However, the next natural question is this: \textit{is that all that determines the internal structures?} In fact, there are much more complications in realistic string-inspired models. For example, let us choose $\omega = -1$ to discuss a dilaton model. However, to be consistent with the experimental results \cite{Bertotti:2003rm}, such a dilaton field should be stabilized by a potential, so to speak $V(\Phi)$. Then this potential will restrict the responses of the Brans-Dicke field. Then will it change the internal structure of a charged black hole so that there appears a Cauchy horizon again? Like this, we can ask whether a certain choice of a string-inspired model (e.g., choosing $\beta$) solely determines the internal structure, or there are lots of choices of parameters that determine the internal structures. This paper will be devoted to answer on this problem.

This paper is organized as follows. In SEC.~\ref{sec:mod}, we discuss a detailed model of the Brans-Dicke theory with a $U(1)$ gauge field, where we further introduce a potential term of the Brans-Dicke field. In SEC.~\ref{sec:mas}, we investigate numerical results for the existence of mass inflation and Cauchy horizons by varying some parameters, especially the coupling parameter $\beta$ and the mass scale of the potential. In SEC.~\ref{sec:dis}, we summarize our results and discuss on possible future issues. In Appendices~A, B, and C, we discussed details of our numerical formalism, initial settings, convergence and consistency checks, and some details of numerical results.

\section{\label{sec:mod}Model for charged black holes}

In this section, we discuss on the details of the model for a Brans-Dicke theory with a $U(1)$ gauge field. In addition, we briefly summarize the previous results of the authors \cite{Hansen:2014rua} and discuss on the motivation of this paper.

\subsection{Brans-Dicke theory with a $U(1)$ gauge field}

The prototype action of the Brans-Dicke theory with a $U(1)$ gauge field becomes ($c=G=\hbar=1$)
\begin{eqnarray}\label{eq:BDscalar}
S = \int \sqrt{-g}d^{4}x \left[\frac{1}{16\pi} \left( \Phi R - \frac{\omega}{\Phi}\Phi_{;\mu}\Phi_{;\nu}g^{\mu\nu} - V(\Phi)\right) + \Phi^{\beta} \mathcal{L}^{\mathrm{EM}} \right],
\end{eqnarray}
where
\begin{eqnarray}
\mathcal{L}^{\mathrm{EM}} \equiv - \frac{1}{2}\left(\phi_{;\mu}+ieA_{\mu}\phi \right)g^{\mu\nu}\left(\overline{\phi}_{;\nu}-ieA_{\nu}\overline{\phi}\right)-\frac{1}{16\pi}F_{\mu\nu}F^{\mu\nu},
\end{eqnarray}
$\Phi$ is the Brans-Dicke field, $\omega$ and $\beta$ are free parameters that determine the model, $\phi$ is a complex scalar field with a gauge coupling $e$, $A_{\mu}$ is a gauge field, $F_{\mu\nu} \equiv A_{\nu;\mu}-A_{\mu;\nu}$, and $V(\Phi)$ is the potential of the Brans-Dicke field, where we use the form
\begin{eqnarray}
V(\Phi) = \frac{1}{2} M^{2} \left( \Phi - 1 \right)^{2}
\end{eqnarray}
with a constant $M$.

The Einstein equation becomes as follows:
\begin{eqnarray}\label{eq:Einstein}
G_{\mu\nu} = 8 \pi \left( T^{\mathrm{BD}}_{\mu\nu} + \Phi^{\beta-1} T^{\mathrm{C}}_{\mu\nu} \right),
\end{eqnarray}
where the Brans-Dicke part and the matter part of the energy-momentum tensors are
\begin{eqnarray}\label{eq:T_BD}
T^{\mathrm{BD}}_{\mu\nu} &=& \frac{1}{8\pi \Phi} \left(-g_{\mu\nu}\Phi_{;\rho \sigma}g^{\rho\sigma}+\Phi_{;\mu\nu}\right)
+ \frac{\omega}{8\pi \Phi^{2}} \left(\Phi_{;\mu}\Phi_{;\nu}-\frac{1}{2}g_{\mu\nu}\Phi_{;\rho}\Phi_{;\sigma}g^{\rho\sigma}\right) - \frac{V(\Phi)}{16 \pi \Phi} g_{\mu\nu},\\
\label{eq:T_C}
T^{\mathrm{C}}_{\mu\nu} &=& \frac{1}{2}\left(\phi_{;\mu}\overline{\phi}_{;\nu}+\overline{\phi}_{;\mu}\phi_{;\nu}\right)
+\frac{1}{2}\left(-\phi_{;\mu}ieA_{\nu}\overline{\phi}+\overline{\phi}_{;\nu}ieA_{\mu}\phi+\overline{\phi}_{;\mu}ieA_{\nu}\phi-\phi_{;\nu}ieA_{\mu}\overline{\phi}\right)
\nonumber \\
&& {}+\frac{1}{4\pi}F_{\mu \rho}{F_{\nu}}^{\rho}+e^{2}A_{\mu}A_{\nu}\phi\overline{\phi}+\mathcal{L}^{\mathrm{EM}}g_{\mu \nu}.
\end{eqnarray}
In addition, the field equations for matter fields are as follows:
\begin{eqnarray}
\label{eq:Phi}0 &=& \Phi_{;\mu\nu}g^{\mu\nu}-\frac{8\pi \Phi^{\beta}}{3+2\omega} \left(T^{\mathrm{C}} - 2\beta \mathcal{L}^{\mathrm{EM}} \right) - \frac{1}{3+2\omega} \left( \Phi V'(\Phi) - 2 V(\Phi) \right), \\
\label{eq:phi}0 &=& \phi_{;\mu\nu}g^{\mu\nu}+ieA^{\mu}\left(2\phi_{;\mu}+ieA_{\mu}\phi\right)+ieA_{\mu;\nu}g^{\mu\nu}\phi + \frac{\beta}{\Phi}\Phi_{;\mu}\left(\phi_{;\nu}+ieA_{\nu}\phi\right)g^{\mu\nu},
\\
\label{eq:A}0 &=& \frac{1}{2\pi}\left({F^{\nu}}_{\mu;\nu} + \frac{\beta}{\Phi}{F^{\nu}}_{\mu}\Phi_{;\nu} \right) -ie\phi\left(\overline{\phi}_{;\mu}-ieA_{\mu}\overline{\phi}\right)+ie\overline{\phi}\left(\phi_{;\mu}+ieA_{\mu}\phi\right),
\end{eqnarray}
where $T^{\mathrm{C}} \equiv {T^{\mathrm{C}}}^{\mu}_{\;\mu}$.

\begin{figure}
\begin{center}
\includegraphics[scale=0.8]{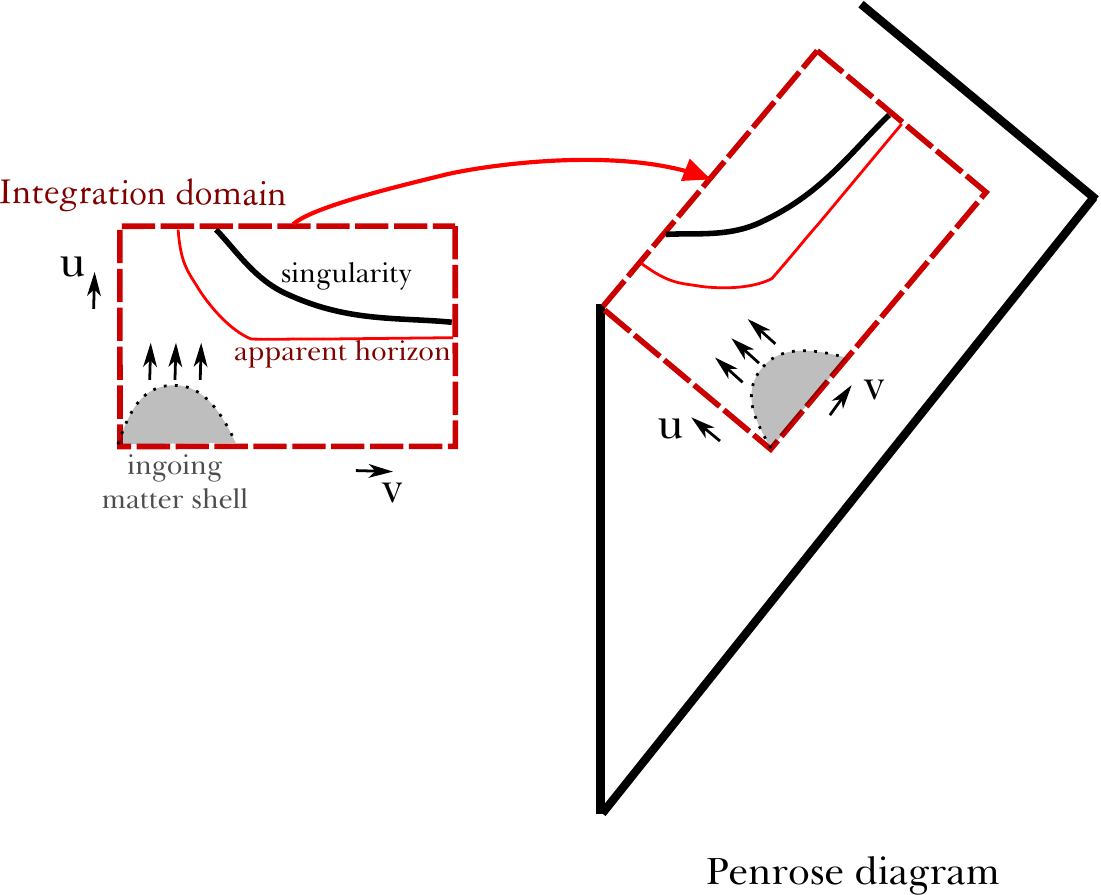}
\caption{\label{fig:domain}Conceptual interpretation of the results of double-null simulations. Left is the numerical integration domain. By tilting $45$ degrees, we can interpret this as a Penrose diagram.}
\end{center}
\end{figure}

In the double-null formalism, we use the double-null coordinates
\begin{eqnarray}\label{eq:doublenull}
ds^{2} = -\alpha^{2}(u,v) du dv + r^{2}(u,v) d\Omega^{2}
\end{eqnarray}
and present every equations using this metric ansatz. Here, we assume the spherical symmetry, $u$ is the retarded time, $v$ is the advanced time, $d\Omega^{2} = d\theta^{2} + \sin^{2} \theta d\varphi^{2}$, where $\theta$ and $\varphi$ are angular coordinates. The detailed formulations and assignments of initial conditions are discussed in Appendices A and B. We need to assign the initial conditions for the in-going and out-going null slices. Especially we prepare a condition for a collapsing shell along the initial out-going null slice. After we get a two dimensional data by solving numerical simulations as functions of $u$ and $v$, by tilting $45$ degrees, we can interpret them as a Penrose diagram (FIG.~\ref{fig:domain}).

\subsection{Summary of previous results}

In the previous paper of the authors \cite{Hansen:2014rua}, we investigated causal structures and responses of the Brans-Dicke field of string-inspired models. One of the crucial point is the dynamics of the Brans-Dicke field that satisfies ($M=0$)
\begin{eqnarray}
\nabla^{2} \Phi + \frac{(\beta-1)}{3+2\omega} \times \mathcal{K} + \frac{\beta}{3+2\omega} \times \mathcal{Q} = 0,
\end{eqnarray}
where
\begin{eqnarray}
\mathcal{K} &\equiv& 8\pi \Phi^{\beta} \left(\frac{w\bar{z} + z\bar{w} + iea (\bar{z}s-z\bar{s})}{2\pi \alpha^{2}}\right),\\
\mathcal{Q} &\equiv& 8\pi \Phi^{\beta} \frac{q^{2}}{4 \pi r^{4}}.
\end{eqnarray}
Here, $\mathcal{Q}$ can be interpreted as a charge term that remains after gravitational collapse, while $\mathcal{K}$ is a kinetic term of the matter field that only contribute during a gravitational collapse and disappear as time goes on. Therefore, if we assume $\omega > -3/2$, then
\begin{itemize}
\item[--] \textit{During} gravitational collapse, via the $\mathcal{K}$ term,
\begin{itemize}
\item[--] if $\beta > 1$, then the Brans-Dicke field towards a weak coupling limit,
\item[--] if $\beta < 1$, then the Brans-Dicke field towards a strong coupling limit,
\end{itemize}
\item[--] \textit{After} gravitational collapse, via the $\mathcal{Q}$ term,
\begin{itemize}
\item[--] if $\beta > 0$, then the Brans-Dicke field towards a weak coupling limit,
\item[--] if $\beta < 0$, then the Brans-Dicke field towards a strong coupling limit.
\end{itemize}
\end{itemize}
Therefore, for string-inspired models with $0 < \beta \leq 1$, after gravitational collapse, the Brans-Dicke field towards the weak coupling limit; and, in these cases, as we will repeat again in the next section, there is no Cauchy horizon. On the other hand, if there is a tendency that the Brans-Dicke field is not biased ($\beta = 0$), there exists a Cauchy horizon.

\subsection{Task of this paper}

Then can we conclude that, for Type I ($\beta = 1/2$) or Heterotic model ($\beta = 1$), always there is no Cauchy horizon? However, in general the situations are complicated. In this paper, we vary more parameters.
\begin{description}
\item[-- $\beta$ dependence:] Even though there is no good corresponding model in string theory, as a theoretical consideration, we can investigate for $\beta < 0$ or $\beta > 1$ limits. For example, if $\beta < 0$, then during and after gravitational collapses, the Brans-Dicke field will be biased to a strong coupling limit. Then can there be a Cauchy horizon again? In other words, does the tendency of the bias of the Brans-Dicke field solely determine the existence of the Cauchy horizon?
\item[-- Potential dependence:] In general the dilaton field should be stabilized by a potential. This gives a correction term to the equation so that
\begin{eqnarray}
\nabla^{2} \Phi + \frac{(\beta-1)}{3+2\omega} \times \mathcal{K} + \frac{\beta}{3+2\omega} \times \mathcal{Q} - \frac{1}{3+2\omega} \times \mathcal{P} = 0,
\end{eqnarray}
where
\begin{eqnarray}
\mathcal{P} \equiv \Phi V' - 2 V
\end{eqnarray}
is a contribution from the potential. For any potential, around the local minimum, it can be well approximated by a quadratic form $V = M^{2}(\Phi-1)^{2}/2$, and hence $\mathcal{P} = M^{2} ( \Phi - 1 )$. Then what is the relation with the existence of a Cauchy horizon and the mass parameter $M$?
\end{description}

\section{\label{sec:mas}Mass inflation and existence of Cauchy horizons}

In Einstein gravity, mass inflation is a general phenomena that appears inside a charged black hole \cite{Poisson:1990eh}. For four dimensional cases, a static charged black hole metric is
\begin{eqnarray}
ds^{2} = - \left( 1 - \frac{2m}{r} + \frac{q^{2}}{r^{2}} \right) dt^{2} + \left( 1 - \frac{2m}{r} + \frac{q^{2}}{r^{2}} \right)^{-1} dr^{2} + r^{2} d\Omega^{2}.
\end{eqnarray}
In this case, there are two horizons, one is an outer event horizon and the other is an inner Cauchy horizon. If there is a pulse of any matter or energy along the in-going null direction, then it can be well approximated by $T_{\mu\nu} \simeq \partial_{\mu}v \partial_{\nu}v F(v)/r^{2}$, where $F(v)$ is the luminosity function that is proportional to $v^{-p}$ with a constant $p$ for astrophysical matters. Now if an observer falls into a black hole and approaches the Cauchy horizon following the increasing $v$ direction, then the observer measures the energy density $\rho = T_{\mu\nu}l^{\mu}l^{\nu} \simeq e^{2\kappa_{-}v}$, where $l_{\mu}$ is a tangent vector of the observer and $\kappa_{-}$ is the surface gravity at the inner horizon. Therefore, the local energy density should exponentially increases as an observer approaches to the inner Cauchy horizons; hence, as a back-reaction, relating curvature quantities should increase exponentially around the inner horizon.

\begin{figure}
\begin{center}
\includegraphics[scale=0.14]{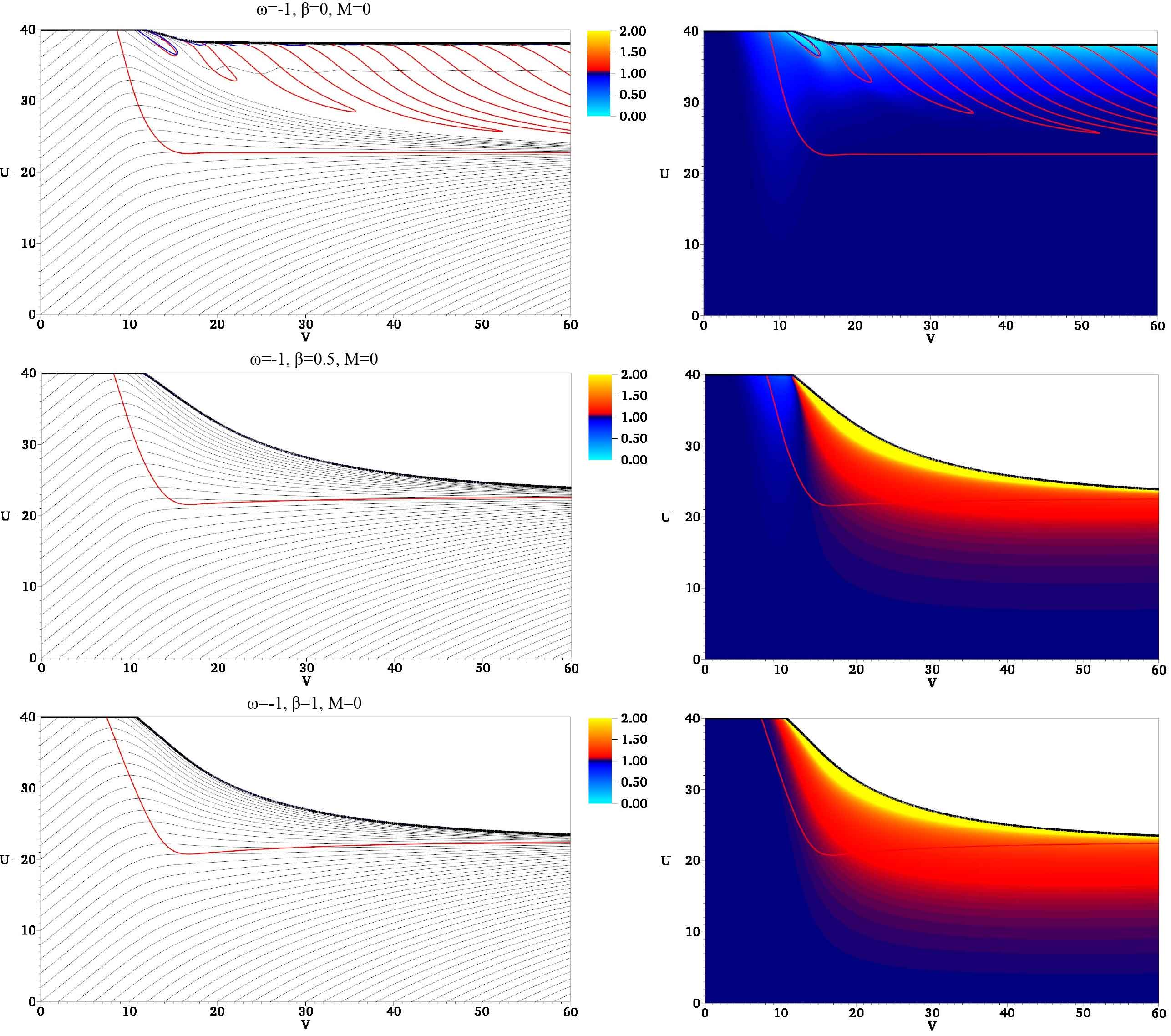}
\caption{\label{fig:M=0}$r$ (left) and $\Phi$ (right) for $\beta = 0$ (top), $0.5$ (middle), $1$ (bottom) with $M = 0$. Thick black curves denote a space-like singularity ($r=0$), thin black curves denote $r = \mathrm{const.}$ contours, red curves denote $r_{,v}=0$ apparent horizons, and blue curves denote $r_{,u}=0$ horizons.}
\end{center}
\end{figure}

In general, since the curvature quantities are increased seriously, the back-reaction to the metric should be very large and hence we need to rely on numerical calculations to see the detailed internal structures \cite{Yeom3}. If there are some contributions from higher curvature corrections \cite{Yeom5} or a black geometry has a different topology \cite{Hansen:2013vha}, then the tendency of mass inflation can be changed.

In this section, we see more details on the relations of the Brans-Dicke field and the existence of a Cauchy horizon. The first question is this: does the biased direction of the Brans-Dicke field determine the existence of the Cauchy horizon?

\subsection{String inspired models with the $M=0$ limit}

First, we report on the case for $\beta = 0, 0.5, 1$ with $M=0$ (FIG.~\ref{fig:M=0}) where the other parameters are fixed and commented in Appendix B. As we commented, if $\beta > 0$, after the formation of the black hole, the Brans-Dicke field is forced toward the weak coupling region (i.e., $\Phi > 1$), and in the end, there becomes no inner Cauchy horizon.

\subsection{Variation of parameters}

\subsubsection{$\beta < 0$ limit}

Now let us vary the parameter $\beta$ less than zero. Then as we discussed, we will surely see the bias of the Brans-Dicke field toward the strong coupling limit after a formation of a black hole. Then can this guarantee the existence of a Cauchy horizon?

\begin{figure}
\begin{center}
\includegraphics[scale=0.14]{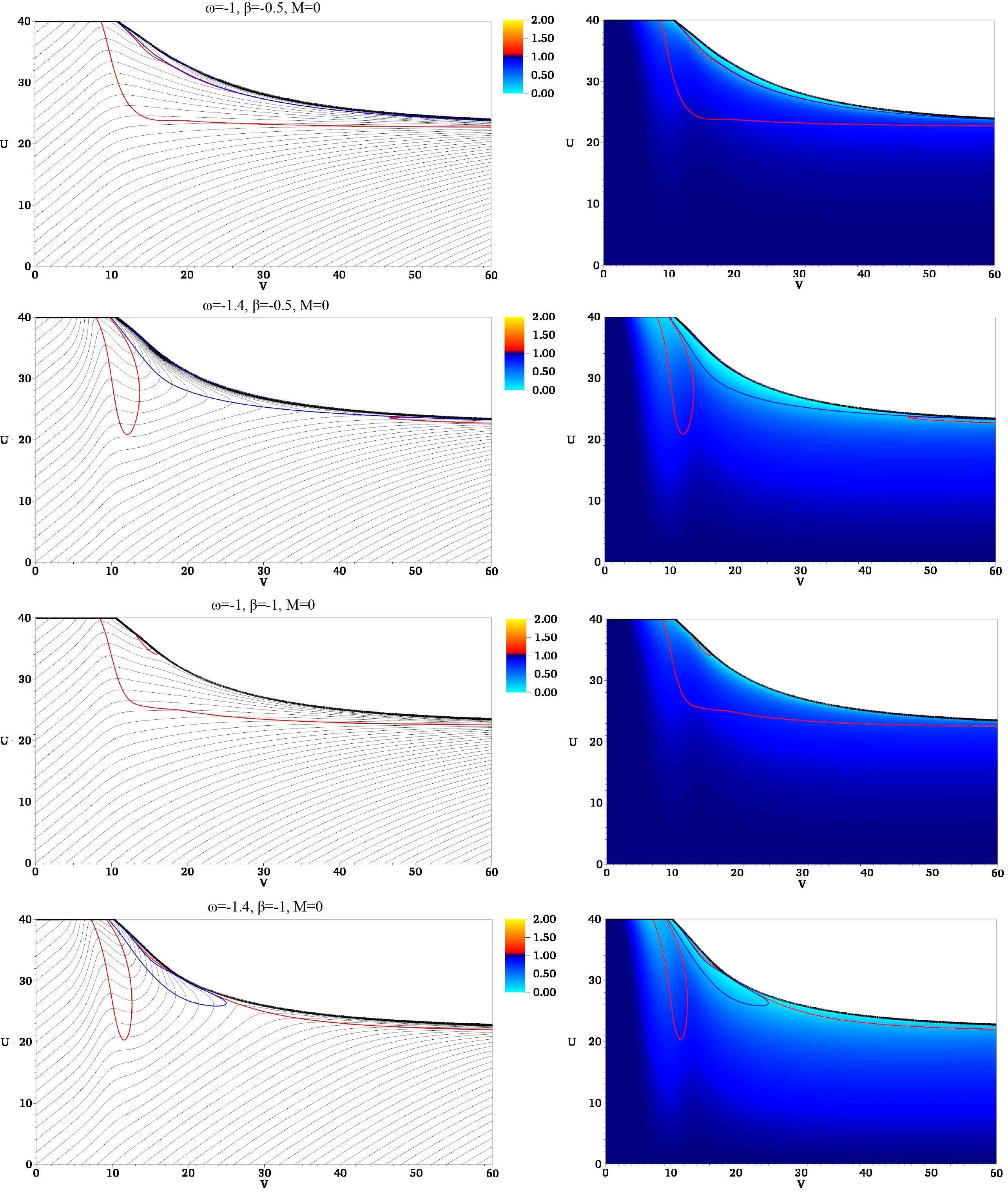}
\caption{\label{fig:Varyingbeta}$r$ (left) and $\Phi$ (right) by varying $\beta = -0.5$ and $\beta = -1$, while we choose $M=0$. In addition, for a comparison, we also tested not only $\omega = -1$, but also $\omega = -1.4$. Here, the thick black curves are singularities in terms of the Einstein frame.}
\end{center}
\end{figure}

To check this, we study the cases of $\beta = -1, -0.5$ with $M = 0$ (FIG.~\ref{fig:Varyingbeta}). Although we are interested in the case $\omega=-1$ (dilaton limit), to amplify the dynamics of the Brans-Dicke field, we also investigate the cases of $\omega = -1.4$.

\begin{figure}
\begin{center}
\includegraphics[scale=0.5]{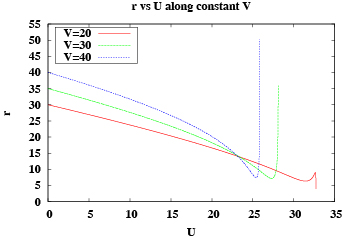}
\includegraphics[scale=0.5]{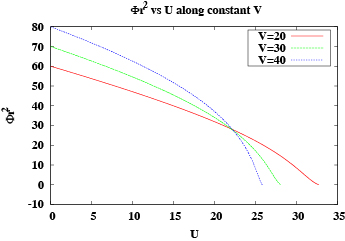}
\caption{\label{fig:rvsU}$r$ (left) and $\Phi r^{2}$ (right) for $\beta = -0.5$ and $\omega = -1$ case. Although $r$ increases to a large value for an in-going null observer, the physical areal radius in the Einstein frame goes to zero (hence, towards a singularity).}
\end{center}
\end{figure}

These results are impressive. Let us summarize important points:
\begin{itemize}
\item[--] As we can see the Brans-Dicke field, during and after the gravitational collapse, the Brans-Dicke field is biased toward the strong coupling regime.
\item[--] For some cases (first, second, and fourth in FIG.~\ref{fig:Varyingbeta}), inside the event horizon, there exists not only $r_{,v}=0$ horizon but also $r_{,u} = 0$ horizon. This means that the in-going observer will see an increasing areal radius (left of FIG.~\ref{fig:rvsU}). This is a kind of wormhole inside the black hole. This is extremely surprising, but we can understand this. In the Einstein frame, the spherical area is proportional to $\propto \Phi r^{2}$, where $\Phi$ and $r$ are calculated by the Jordan frame. Therefore, in this case, as an observer falls into the black hole, the observer experiences $r \rightarrow \infty$ and $\Phi r^{2} \rightarrow 0$ (right of FIG.~\ref{fig:rvsU}). Therefore, in the Jordan frame, the thick black curves in FIG.~\ref{fig:Varyingbeta} may not be $r=0$, but we should regard the region as the singularity.
\item[--] If we understand this, then in fact, all of these results shows that there is no Cauchy horizon and mass inflation. The causal structure is the same as that of the neutral black hole and the cases of $\beta > 0$.
\end{itemize}
Therefore, we can conclude that the existence of the Cauchy horizon and mass inflation is \textit{not} determined by the direction of the bias of the Brans-Dicke field. After a charged black hole is formed, \textit{at once it is biased} toward strong or weak couping regime, then there is no Cauchy horizon. Therefore, this strongly indicates that the existence of a Cauchy horizon and mass inflation is related the fact that there should be \textit{no} bias of the Brans-Dicke field; e.g., there should be no hair of the Brans-Dicke field outside the horizon. Perhaps, the existence of a Cauchy horizon and mass inflation is the property of the pure Einstein gravity, while this can be destroyed by introducing complicated couplings that creates Brans-Dicke hairs.


\begin{figure}
\begin{center}
\includegraphics[scale=0.14]{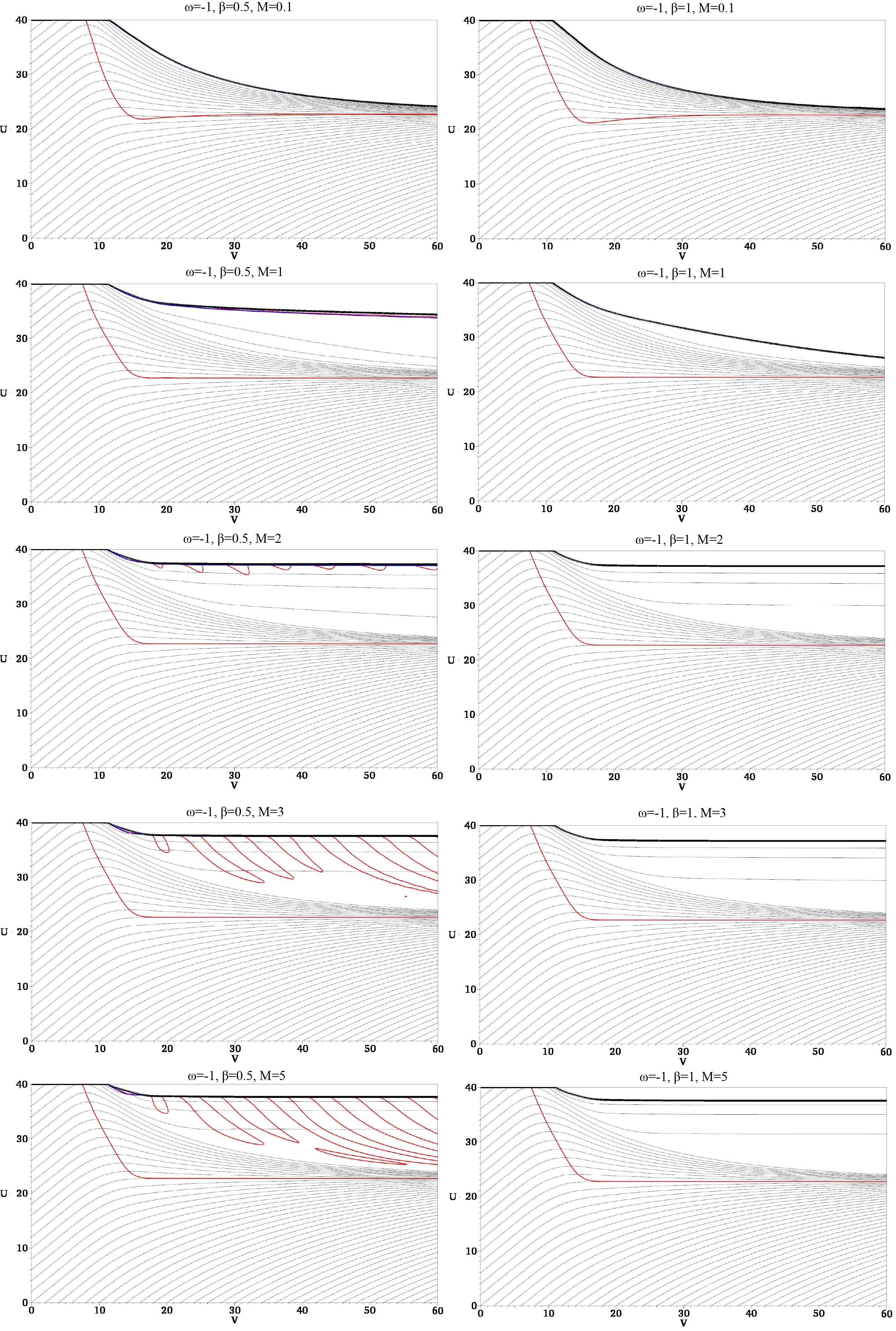}
\caption{\label{fig:VaryingM}$r$ by varying $M = 0.1$, $1$, $2$, $3$, and $5$ (from top to bottom) for $\beta = 0.5$ (left) and $1$ (right).}
\end{center}
\end{figure}

\begin{figure}
\begin{center}
\includegraphics[scale=0.14]{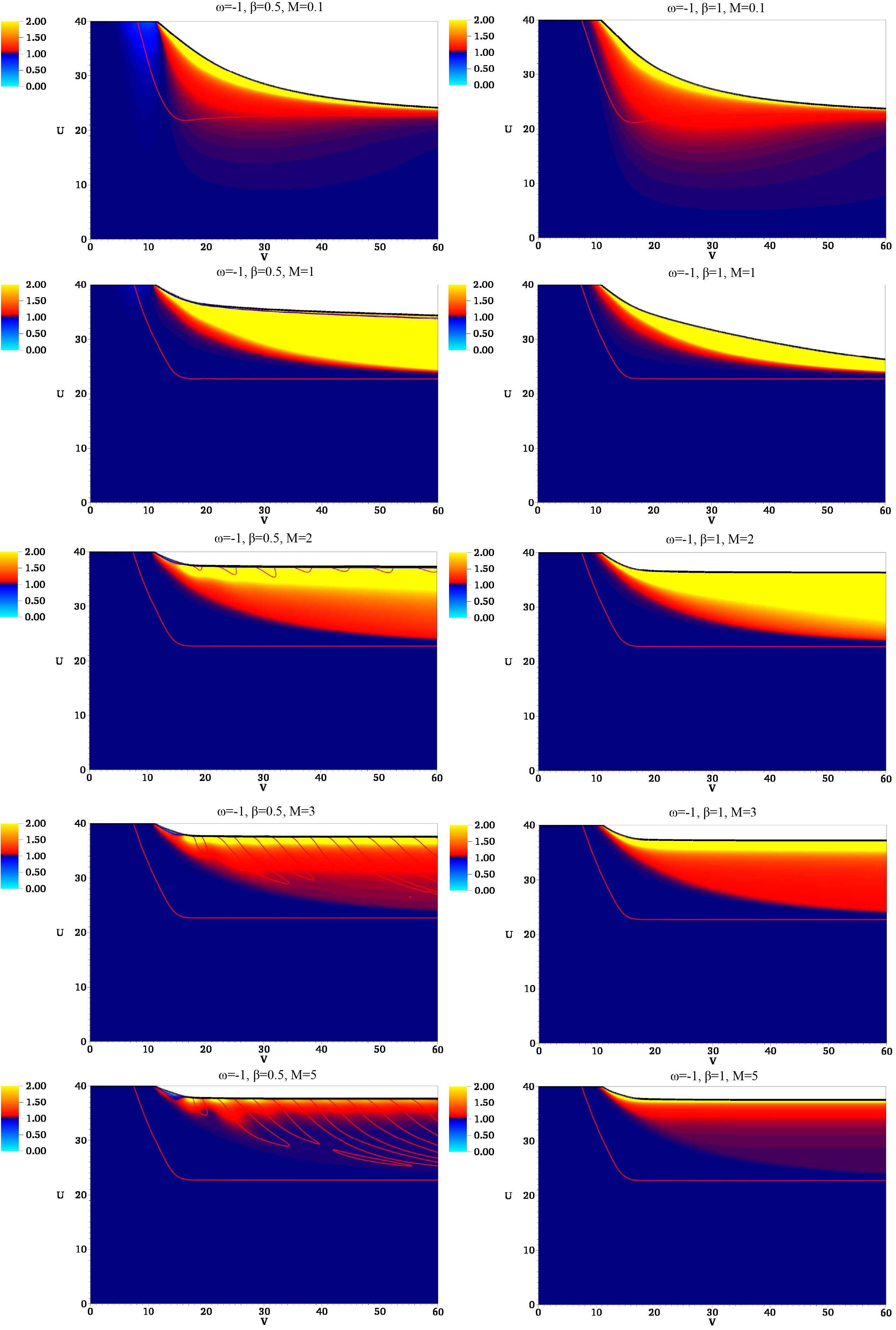}
\caption{\label{fig:VaryingM_Phi}$\Phi$ by varying $M = 0.1$, $1$, $2$, $3$, and $5$ (from top to bottom) for $\beta = 0.5$ (left) and $1$ (right).}
\end{center}
\end{figure}

\begin{figure}
\begin{center}
\includegraphics[scale=0.14]{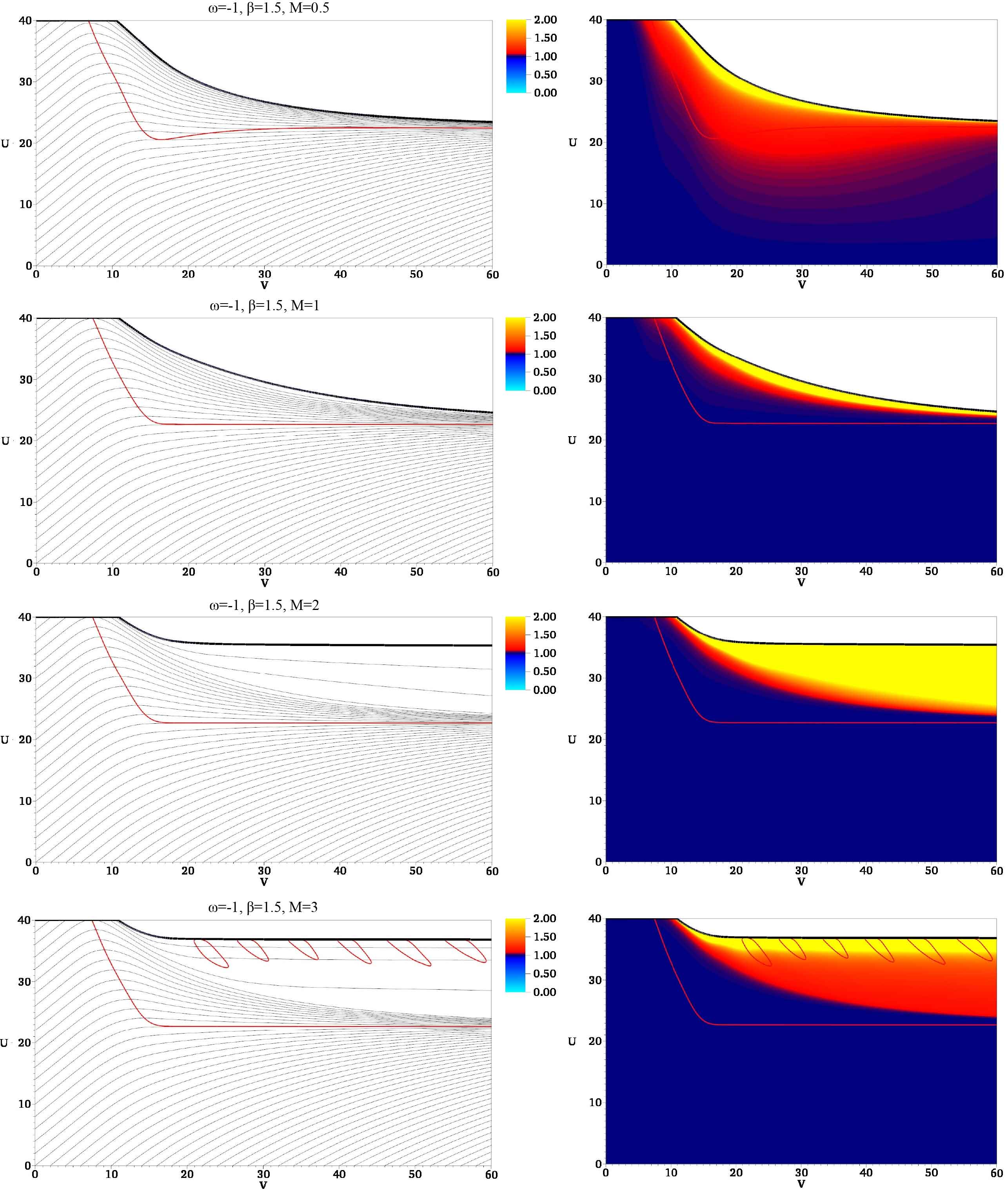}
\caption{\label{fig:VaryingbetaM}$r$ (left) and $\Phi$ (right) for $\beta = 1.5$ by varying $M = 0.1$, $1$, $2$, and $3$ (from top to bottom).}
\end{center}
\end{figure}

\subsubsection{$M>0$ limit}

Then unless $\beta = 0$, are there any hope to recover a Cauchy horizon? By introducing a potential of the Brans-Dicke field, we can adjust the Brans-Dicke hair, and this can be helpful to re-create a Cauchy horizon.

Comparing with FIG.~\ref{fig:M=0}, in FIG.~\ref{fig:VaryingM}, we can vary $M$ from $0.1$ to $5$. This shows that as we increase the effect of $M$, the causal structure is changed and the inner Cauchy horizon appears again as the case of the usual charged black holes with Einstein gravity. If we compare with $\Phi$ (FIG.~\ref{fig:VaryingM_Phi}), then we can see clear dependence. Because of the choice of $\beta$, after a formation of a black hole, the Brans-Dicke field is biased toward a weak coupling limit, and this is the same as that of the case of $M=0$. However, because of the mass parameter, the sensitivity of the Brans-Dicke field is limited. Therefore, if the mass parameter is small enough (e.g., $M = 0.1$) so that the Brans-Dicke field is biased to the weak coupling limit even \textit{outside} the outer apparent horizon, then the effects of charge is already screened and hence the internal structure does not have a Cauchy horizon. However, as $M$ increases, e.g., for the case of $M = 1$ or $2$, although the Brans-Dicke field inside the horizon is biased toward the weak coupling limit, it is no more biased in terms of the outside the outer apparent horizon. If we choose much larger value of $M$ (e.g., $M = 5$), then even the inside the outer apparent horizon, the bias of the Brans-Dicke field is limited and decreases.

In addition, as a check, we observe the case when $\beta > 1$ and varying $M$ in FIG.~\ref{fig:VaryingbetaM}. As we expected, for example if $\beta = 1.5$, during and after the gravitational collapse, the Brans-Dicke field should be biased to weak coupling direction, and hence there should be no Cauchy horizon. However, by increasing $M$, the Brans-Dicke hair is controlled and eventually there appears a Cauchy horizon and mass inflation.

\section{\label{sec:dis}Discussion}

In this paper, we investigated dynamics of string-inspired charged black holes. In our previous paper \cite{Hansen:2014rua}, we already investigated the causal structures and responses of the Brans-Dicke field. Regarding the causal structure, the main points were as follows: (1) if $\beta > 0$, then after the gravitational collapse, there appears a Brans-Dicke hair that is biased toward a weak coupling limit and (2) unless $\beta = 0$, there is no Cauchy horizon and no mass inflation inside a charged black hole.

In this paper, we investigated the physical origin of the existence or absence of mass inflation and the Cauchy horizon. What we have observed are as follows:
\begin{itemize}
\item[--] If there is no potential of the Brans-Dicke field, unless $\beta = 0$, after the gravitational collapse, the Brans-Dicke hair will be formed and biased either toward a weak coupling limit (if $\beta > 0$) or a strong coupling limit (if $\beta < 0$). For all cases, there is no Cauchy horizon and mass inflation; hence, the existence/absence of a Brans-Dicke hair is related to the existence/absence of mass inflation and Cauchy horizon inside a charged black hole.
\item[--] Even though $\beta \neq 0$, by introducing a potential of the Brans-Dicke field, we can reproduce a mass inflation and a Cauchy horizon. In other words, if the Brans-Dicke field is well trapped by a potential, then even though the Brans-Dicke field is affected by charges of the black hole, we can still see a mass inflation and a Cauchy horizon.
\end{itemize}
Therefore, we can qualitatively conclude that the existence of mass inflation is related to the existence of the Brans-Dicke hair. In other words, we may further say that (within some limited conditions) the existence of mass inflation and a Cauchy horizon inside a charged black hole is a consequence of no scalar hair.

In string theory, the dilaton field should be well trapped by a potential; unless it is unrealistic. Therefore, it may be sound to conclude that there may exist mass inflation inside realistic charged black holes in our universe. On the other hand, if the trap of the dilaton field is not deep enough and hence a kind of scalar hair is detectable by an experiment, then we may conclude that the black hole just have a space-like singularity without a Cauchy horizon.

It may be interesting that we can know the internal information of a black hole by observing the dilaton behaviors outside the horizon. However, our descriptions are still qualitative. To say more details, analytic modeling and calculations would be required to give a detailed connections between the dilaton hair shape and the internal structures. In addition, the dependence or sensitivity by varying another parameters, for examples space-dimensions, topology, or the background cosmological constant, can be an interesting future topic. We remain them for a future research topic.

\newpage

\section*{Acknowledgments}
DY was supported by Leung Center for Cosmology and Particle Astrophysics (LeCosPA) of National Taiwan University (103R4000). JH was supported in part by the programs of the Construction and Operation for Large-scale Science Data Center at KISTI (K-14-L01-C06-S01), the Global Hub for Experiment Data of Basic Science by NRF (N-14-NM-IR06), and the APCTP Topical Research Program.

\section*{Appendix A. Implementation to double-null formalism}

We use the double-null coordinates
\begin{eqnarray}
ds^{2} = -\alpha^{2}(u,v) du dv + r^{2}(u,v) d\Omega^{2}
\end{eqnarray}
and present every equations using this metric ansatz. To simplify and present every equations by first order differentials, we define the following variables \cite{Yeom1,Yeom3}: The metric function $\alpha$, the radial function $r$, the Brans-Dicke field $\Phi$ and a complex scalar field $s \equiv \sqrt{4\pi} \phi$, and define
\begin{eqnarray}\label{eq:conventions}
h \equiv \frac{\alpha_{,u}}{\alpha},\quad d \equiv \frac{\alpha_{,v}}{\alpha},\quad f \equiv r_{,u},\quad g \equiv r_{,v},\quad W \equiv \Phi_{,u},\quad Z \equiv \Phi_{,v}, \quad w \equiv s_{,u},\quad z \equiv s_{,v}.
\end{eqnarray}
Using this, the Einstein tensor and the energy-momentum tensor for the Brans-Dicke part and the scalar field part are as follows:
\begin{eqnarray}
\label{eq:Guu}G_{uu} &=& -\frac{2}{r} \left(f_{,u}-2fh \right),\\
\label{eq:Guv}G_{uv} &=& \frac{1}{2r^{2}} \left( 4 rf_{,v} + \alpha^{2} + 4fg \right),\\
\label{eq:Gvv}G_{vv} &=& -\frac{2}{r} \left(g_{,v}-2gd \right),\\
\label{eq:Gthth}G_{\theta\theta} &=& -4\frac{r^{2}}{\alpha^{2}} \left(d_{,u}+\frac{f_{,v}}{r}\right),\\
\label{eq:TBDuu}T^{\mathrm{BD}}_{uu} &=& \frac{1}{8 \pi \Phi} (W_{,u} - 2hW) + \frac{\omega}{8 \pi \Phi^{2}} W^{2},\\
\label{eq:TBDuv}T^{\mathrm{BD}}_{uv} &=& - \frac{Z_{,u}}{8 \pi \Phi} - \frac{gW+fZ}{4 \pi r \Phi} + \frac{\alpha^{2} V}{32 \pi \Phi},\\
\label{eq:TBDvv}T^{\mathrm{BD}}_{vv} &=& \frac{1}{8 \pi \Phi} (Z_{,v} - 2dZ) + \frac{\omega}{8 \pi \Phi^{2}} Z^{2},\\
\label{eq:TBDthth}T^{\mathrm{BD}}_{\theta\theta} &=& \frac{r^{2}}{2 \pi \alpha^{2} \Phi} Z_{,u} + \frac{r}{4 \pi \alpha^{2} \Phi} (gW+fZ) + \frac{\omega}{4\pi \Phi^{2}} \frac{r^{2}}{\alpha^{2}}WZ - \frac{r^{2}V}{16 \pi \Phi},\\
\label{eq:TSuu}T^{\mathrm{C}}_{uu} &=& \frac{1}{4\pi} \left[ w\overline{w} + iea(\overline{w}s-w\overline{s}) +e^{2}a^{2}s\overline{s} \right],\\
\label{eq:TSuv}T^{\mathrm{C}}_{uv} &=& \frac{{(a_{,v})}^{2}}{4\pi\alpha^{2}},\\
\label{eq:TSvv}T^{\mathrm{C}}_{vv} &=& \frac{1}{4\pi} z\overline{z},\\
\label{eq:TSthth}T^{\mathrm{C}}_{\theta\theta} &=& \frac{r^{2}}{4\pi\alpha^{2}} \left[ (w\overline{z}+z\overline{w}) + iea(\overline{z}s-z\overline{s})+\frac{2{(a_{,v})}^{2}}{\alpha^{2}} \right],
\end{eqnarray}
where $q(u,v) \equiv 2r^{2} a_{,v}/\alpha^{2}$ is interpreted as the charge function.

After a simple calculation, we can represent equations for $\alpha_{,uv}$, $r_{,uv}$, and the field equation for $\Phi$. We define $\widetilde{X}\equiv\Phi^{\beta}X$ for any quantity $X$. Then,
\begin{eqnarray}\label{eq:solved}
\left( \begin{array}{c}
\left(\log \alpha\right)_{,uv} \\
r_{,uv}  \\
\Phi_{,uv} 
\end{array} \right)
= \frac{1}{r^{2}} \left( \begin{array}{ccc}
r^{2} & -r & -\frac{r}{2\Phi} \\
0 & r^{2} & -\frac{r^{2}}{2\Phi} \\
0 & 0 & r
\end{array} \right)
\left( \begin{array}{c}
\mathfrak{A} \\
\mathfrak{B} \\
\mathfrak{C}
\end{array} \right),
\end{eqnarray}
\begin{eqnarray}
\label{eq:E1}r_{,uu} &=& 2fh - \frac{r}{2 \Phi} (W_{,u}-2hW) - \frac{r \omega}{2 \Phi^{2}} W^{2} - \frac{4 \pi r}{\Phi} {\widetilde{T}}^{\mathrm{M}}_{uu},\\
\label{eq:E2}r_{,vv} &=& 2gd - \frac{r}{2 \Phi} (Z_{,v}-2dZ) - \frac{r \omega}{2 \Phi^{2}} Z^{2} - \frac{4 \pi r}{\Phi} {\widetilde{T}}^{\mathrm{M}}_{vv},
\end{eqnarray}
where
\begin{eqnarray}
\label{eq:A}\mathfrak{A} &\equiv& -\frac{2\pi \alpha^{2}}{r^{2}\Phi}\widetilde{T}^{\mathrm{C}}_{\theta\theta} - \frac{1}{2r}\frac{1}{\Phi}(gW+fZ) -\frac{\omega}{2\Phi^{2}}WZ + \frac{\alpha^{2}V}{8\Phi}, \\
\label{eq:B}\mathfrak{B} &\equiv& - \frac{\alpha^{2}}{4r} - \frac{fg}{r} + \frac{4 \pi r}{\Phi}\widetilde{T}^{\mathrm{C}}_{uv} - \frac{1}{\Phi}(gW+fZ) + \frac{r\alpha^{2}}{8\Phi}V, \\
\label{eq:C}\mathfrak{C} &\equiv& - fZ - gW - \frac{2\pi r \alpha^{2}}{3+2\omega} \left( \widetilde{T}^{\mathrm{C}} - 2 \beta {\widetilde{\mathcal{L}}}^{\mathrm{EM}} \right) - \frac{r \alpha^{2}}{4 (3+2\omega)} \left( \Phi V' - 2V \right).
\end{eqnarray}
In addition, matter field equations are
\begin{eqnarray} \label{eq:fieldeqns}
\label{eq:a1}a_{,v} &=& \frac{\alpha ^{2} q}{2 r^{2}}, \\
\label{eq:q1}q_{,v} &=& -\frac{ier^{2}}{2} (\overline{s}z-s\overline{z}) - \beta q \frac{Z}{\Phi}, \\
a_{,vv} &=& \frac{\alpha^{2}}{r^{2}} \left( d - \frac{g}{r} \right)q - \frac{ie\alpha^{2}}{4} \left( z\overline{s}-s\overline{z}\right) - \beta q \frac{\alpha^{2} Z}{2r^{2}\Phi},\\
q_{,u} &=& \frac{ier^{2}}{2} (\overline{s}w-s\overline{w}) - r^{2}e^{2}a s \overline{s} - \beta q \frac{W}{\Phi}, \\
\label{eq:a2}a_{,uv} &=& \frac{\alpha^{2}}{r^{2}} \left( h - \frac{f}{r} \right)q + \frac{ie\alpha^{2}}{4} \left( w\overline{s}-s\overline{w}\right) - \frac{\alpha^{2}}{2}e^{2}as\overline{s} - \beta q \frac{\alpha^{2} W}{2r^{2}\Phi},\\
\label{eq:s}s_{,uv} &=& - \frac{fz}{r} - \frac{gw}{r} - \frac{iearz}{r} - \frac{ieags}{r} - \frac{ie}{4r^{2}}\alpha^{2}qs - \frac{\beta}{2\Phi} \left( Wz +Zw + ies a Z \right).
\end{eqnarray} 
We used numerical integration via a standard 4th order Runge-Kutta method. Consistency and convergence tests are discussed in Appendix~B.

\section*{Appendix B. Boundary conditions and free parameters}

We specify boundary conditions of all variables ($\alpha, r, \Phi, s, a$) on the initial $u=u_{\mathrm{i}}$ and $v=v_{\mathrm{i}}$ null surfaces. For all simulations in this paper, we choose $u_{\mathrm{i}}=v_{\mathrm{i}}=0$ and computational domain to have size $v=[0;60]$ and $u=[0;40]$.

By using the gauge freedom to choose $r$, we choose $r(u,0)_{,u}=r_{u0} < 0$ and $r(0,v)_{,v}=r_{v0} > 0$ such that the function $r$ for an in-going observer decreases and that for an out-going observer increases. In addition,　we fix the effective gravitation constant $G=1/\Phi$ as unity and we set $\Phi(u,0)=\Phi(0,v)=1$. 
\begin{itemize}
\item[--] \textit{In-going null direction}: To assign all functions for the in-going null direction, we simply choose $s(u,0)=0$, $\alpha(u,0)=1$, $q(u,0)=0$, and $a(u,0)=0$, so that the interior is not affected by the gravitational collapse. Since the Misner-Sharpe mass function is
\begin{eqnarray}
m(u,v) \equiv \frac{r}{2}\left( 1+\frac{q^{2}}{r^{2}}+4\frac{r_{,u}r_{,v}}{\alpha^{2}} \right),
\end{eqnarray}
if we give a condition that this vanishes at $u=v=0$, it is convenient to choose $r_{,u}(u,0)=-1/2$ and $r_{,v}(0,v)=1/2$. In addition we further choose $r(0,0)=r_0=20$. In addition, by using the constraint equations (Equation~(\ref{eq:E1})), we can complete the assignment of initial conditions.
\item[--] \textit{Out-going null direction}: We choose a function $s(0,v)$ to induce a collapsing shell. We use 
\begin{eqnarray} \label{s_initial}
s(0,v)= A \sin^{4} \left( \pi \frac{v-v_{\mathrm{i}}}{v_{\mathrm{f}}-v_{\mathrm{i}}} \right) \left[ \cos \left( 2 \pi \frac{v-v_{\mathrm{i}}}{v_{\mathrm{f}}-v_{\mathrm{i}}} \right) + i \cos \left( 2 \pi \frac{v-v_{\mathrm{i}}}{v_{\mathrm{f}}-v_{\mathrm{i}}} -  \pi \delta \right)\right]
\end{eqnarray}
for $v_{\mathrm{i}}\leq v \leq v_{\mathrm{f}}$ and otherwise $s(0,v)=0$. By using this, we integrate Equation~(\ref{eq:E2}) to determine $\alpha(0,v)$ on the $u=0$ surface. Also, at the same time, we integrat Equations~(\ref{eq:a1}) and (\ref{eq:q1}) to determine $q(0,v)$ and $a(0,v)$.
\end{itemize}

Now for a pulse, we choose $v_{\mathrm{f}}=20$, $A = 0.15$, and $\delta = 0.5$, leaving four free parameters $(\omega, \beta, e, M)$, where $\omega$ is the Brans-Dicke coupling parameter, $\beta$ is the coupling between the matter sector and the Brans-Dicke field, $e$ is the gauge coupling, and $M$ is the mass parameter of the potential of the Brans-Dicke field. Note that for each choice of model parameters, we assign meanings as follows \cite{Hansen:2014rua}:
\begin{description}
\item[-- $\omega$]: $0$ ($f(R)$ limit), $-1$ (dilaton limit), $-1.4$ (braneworld limit), $-1.6$ (ghost limit),
\item[-- $\beta$]: $0$ (Type IIA), $0.5$ (Type I), $1$ (Heterotic),
\item[-- $e$]: $0$ (neutral black hole), $0.3$ (charged black hole).
\end{description}
Overall this paper, we fix $e = 0.3$ (only consider charged cases), and vary $\omega$, $\beta$, and $M$.

\begin{figure}
\begin{center}
\includegraphics[scale=0.5]{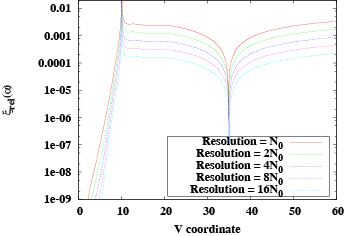}
\includegraphics[scale=0.5]{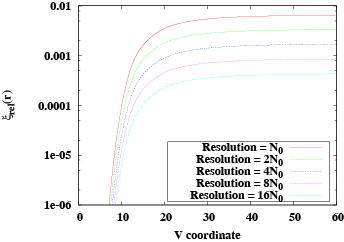}
\includegraphics[scale=0.5]{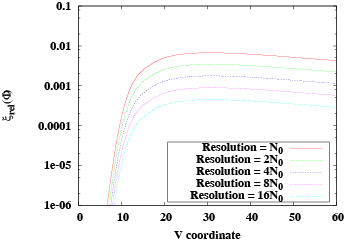}
\includegraphics[scale=0.5]{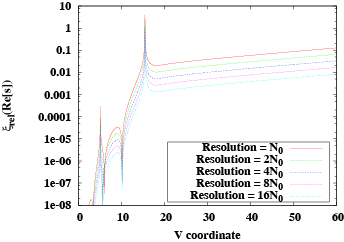}
\includegraphics[scale=0.5]{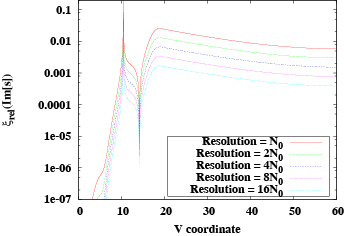}
\includegraphics[scale=0.5]{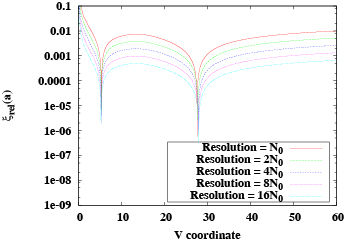}
\includegraphics[scale=0.5]{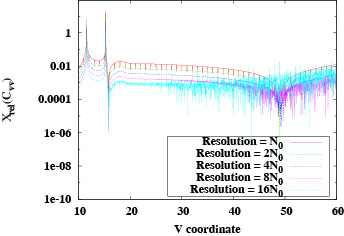}
\includegraphics[scale=0.5]{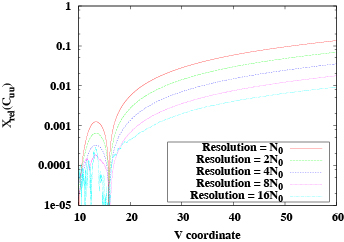}
\caption{\label{fig:convergence}Convergence and consistency checks for along a line $u=20$ for a simulation with parameters: $\omega=-1.4$, $e=0.3$, $\beta=-1.0$, and $M=0$. As the resolution increases, errors are well controlled less than $0.1$~\%.}
\end{center}
\end{figure}

We solved this system by using the same code with the authors' previous paper \cite{Hansen:2014rua}. We repeat a demonstration of convergence and consistency checks: FIG.~\ref{fig:convergence}. Here, we define
\begin{equation}
  \label{eq:xidef} \xi \left( x_N^i \right) \equiv \frac{|x_N^i - x_{2N}^i|}{ |x_{\mathrm{HighRes}}^i |},
\end{equation}
where $x_N^i$ is the $i$-th grid of a dynamic variable $x$ with resolution $N$; $x_{\mathrm{HighRes}}^i$ is that of the highest numerical resolution what we used. In addition, to check constraint equations (Eqs.~(\ref{eq:E1}) and (\ref{eq:E2})), we define
\begin{equation}
  \label{eq:chidef}  \chi \left( C_N^i \right) \equiv \frac{|C_N^i|}{ |G_{\mathrm{HighRes}}^i |},
\end{equation}
where $C_N^i$ is the residual of constraint equations (let us say that $C_{uu}$ or $C_{vv}$) at the $i$-th point for simulation with resolution $N$ and $G_{\mathrm{HighRes}}^i$ is a corresponding Einstein-tensor component ($G_{uu}$ or $G_{vv}$). We can see good convergence and consistency from FIG.~\ref{fig:convergence}: as resolution increases, errors are well controlled and converges less than $0.1$~\%.

\section*{Appendix C. Catalog of metric function and energy-momentum tensors}

\begin{figure}
\begin{center}
\includegraphics[scale=0.14]{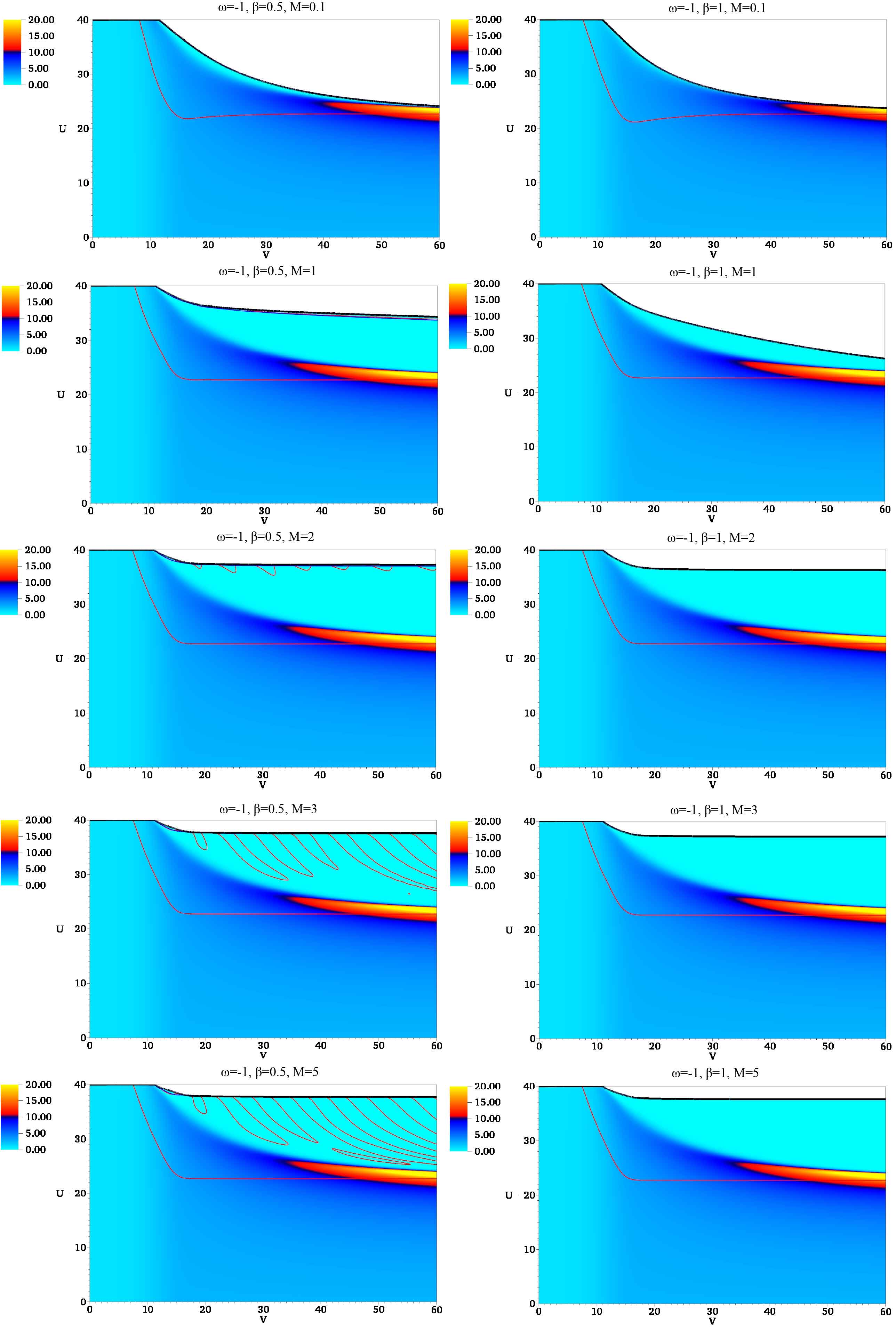}
\caption{\label{fig:alpha_beta}$\alpha^{2}$ by varying $M = 0.1$, $1$, $2$, $3$, and $5$ (from top to bottom) for $\beta = 0.5$ (left) and $1$ (right).}
\end{center}
\end{figure}

\begin{figure}
\begin{center}
\includegraphics[scale=0.14]{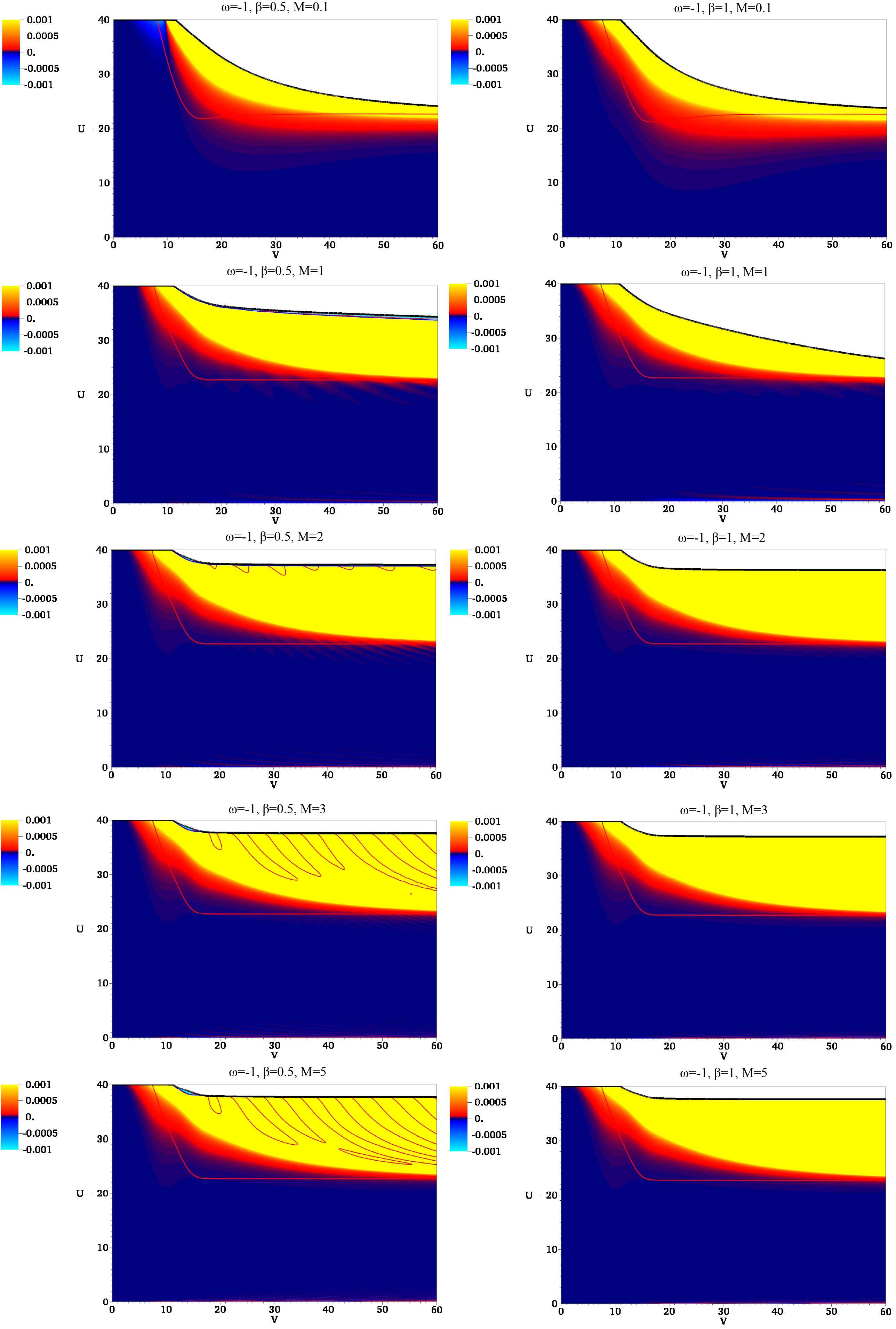}
\caption{\label{fig:Tuu_beta}$T_{uu}$ by varying $M = 0.1$, $1$, $2$, $3$, and $5$ (from top to bottom) for $\beta = 0.5$ (left) and $1$ (right).}
\end{center}
\end{figure}

\begin{figure}
\begin{center}
\includegraphics[scale=0.14]{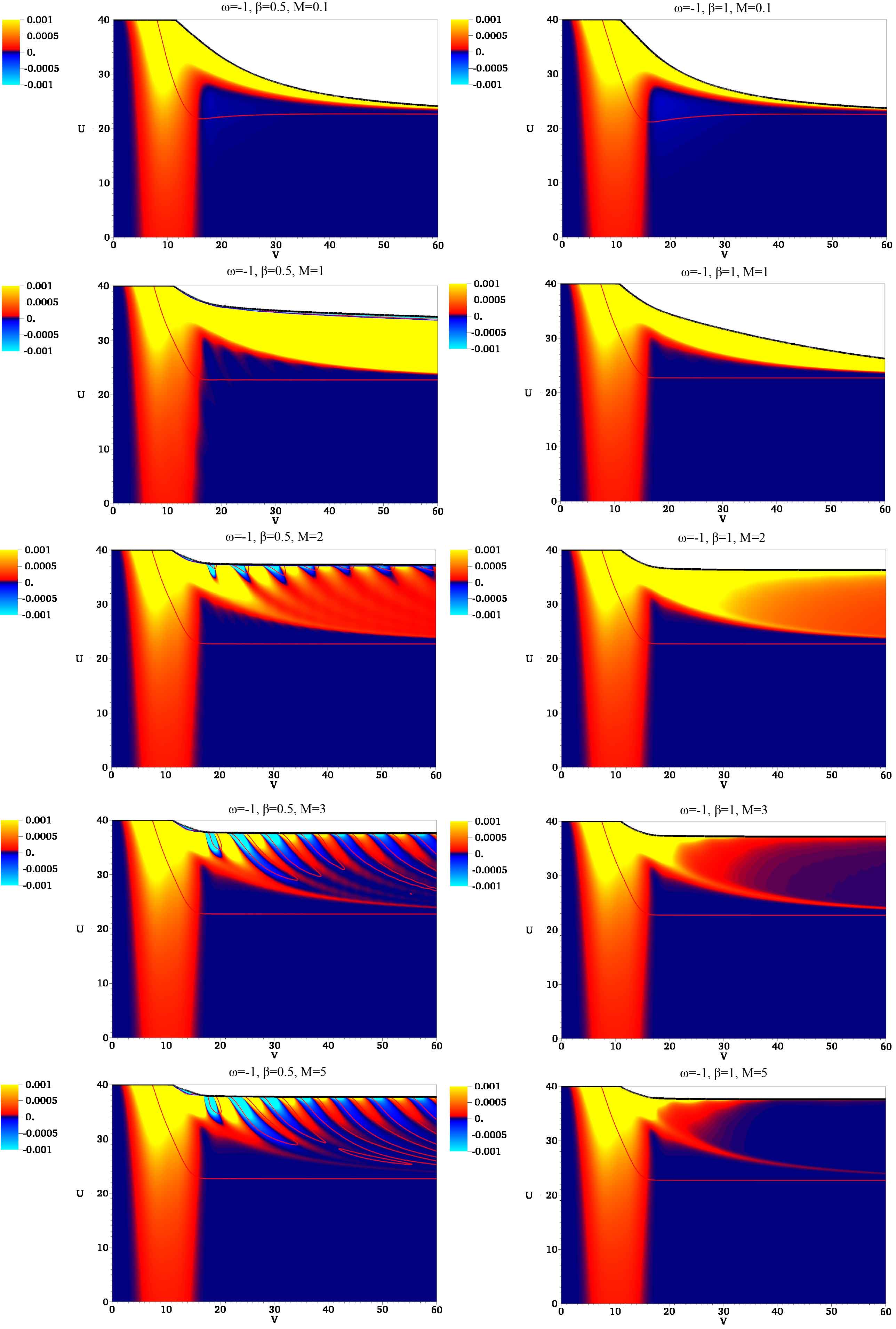}
\caption{\label{fig:Tvv_beta}$T_{vv}$ by varying $M = 0.1$, $1$, $2$, $3$, and $5$ (from top to bottom) for $\beta = 0.5$ (left) and $1$ (right).}
\end{center}
\end{figure}

\begin{figure}
\begin{center}
\includegraphics[scale=0.14]{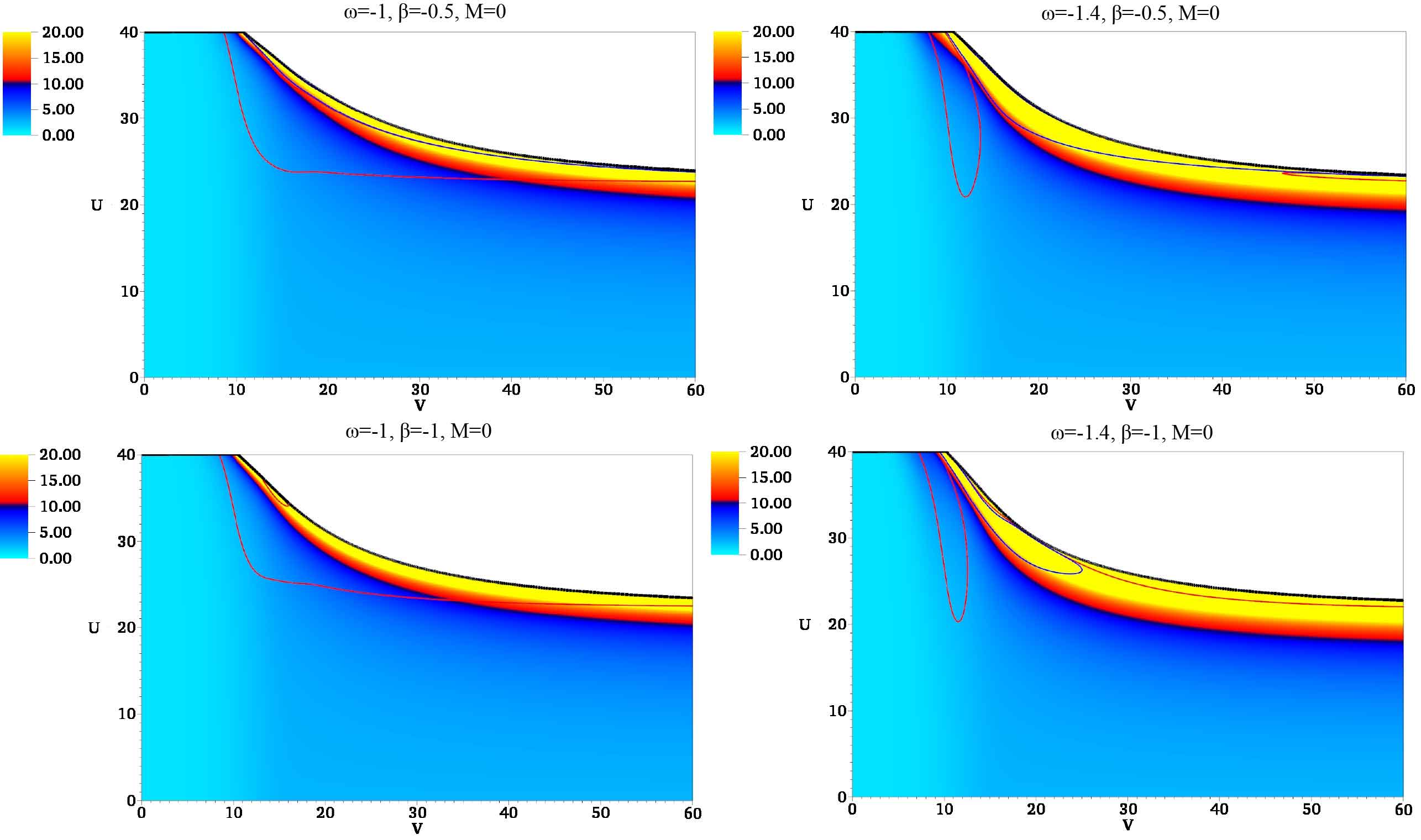}
\caption{\label{fig:alpha_negativebeta}$\alpha^{2}$ by varying $\beta = -0.5$ (upper), $-1$ (lower) and varying $\omega = -1$ (left), $-1.4$ (right).}
\end{center}
\end{figure}

\begin{figure}
\begin{center}
\includegraphics[scale=0.14]{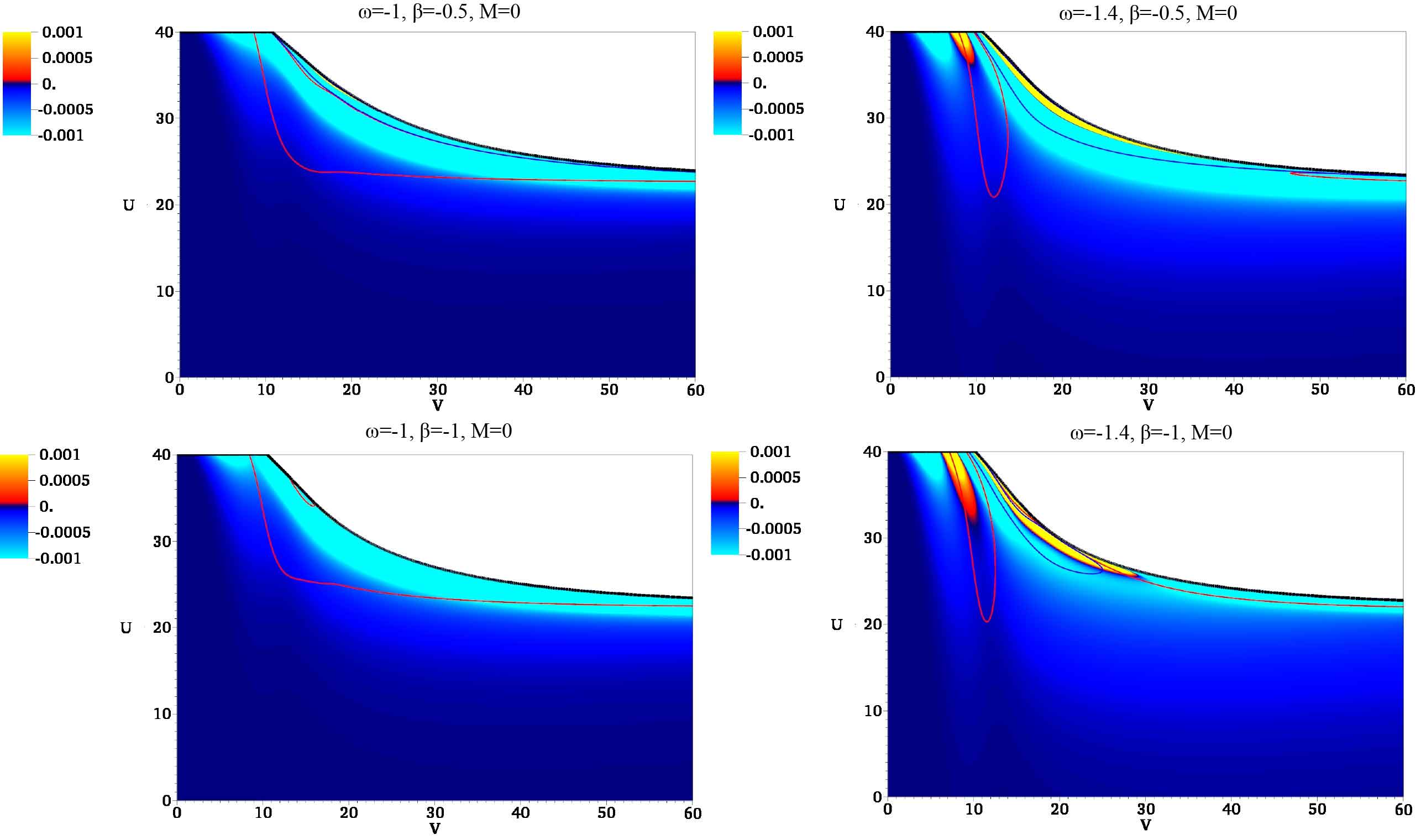}
\caption{\label{fig:Tuu_negativebeta}$T_{uu}$ by varying $\beta = -0.5$ (upper), $-1$ (lower) and varying $\omega = -1$ (left), $-1.4$ (right).}
\end{center}
\end{figure}

\begin{figure}
\begin{center}
\includegraphics[scale=0.14]{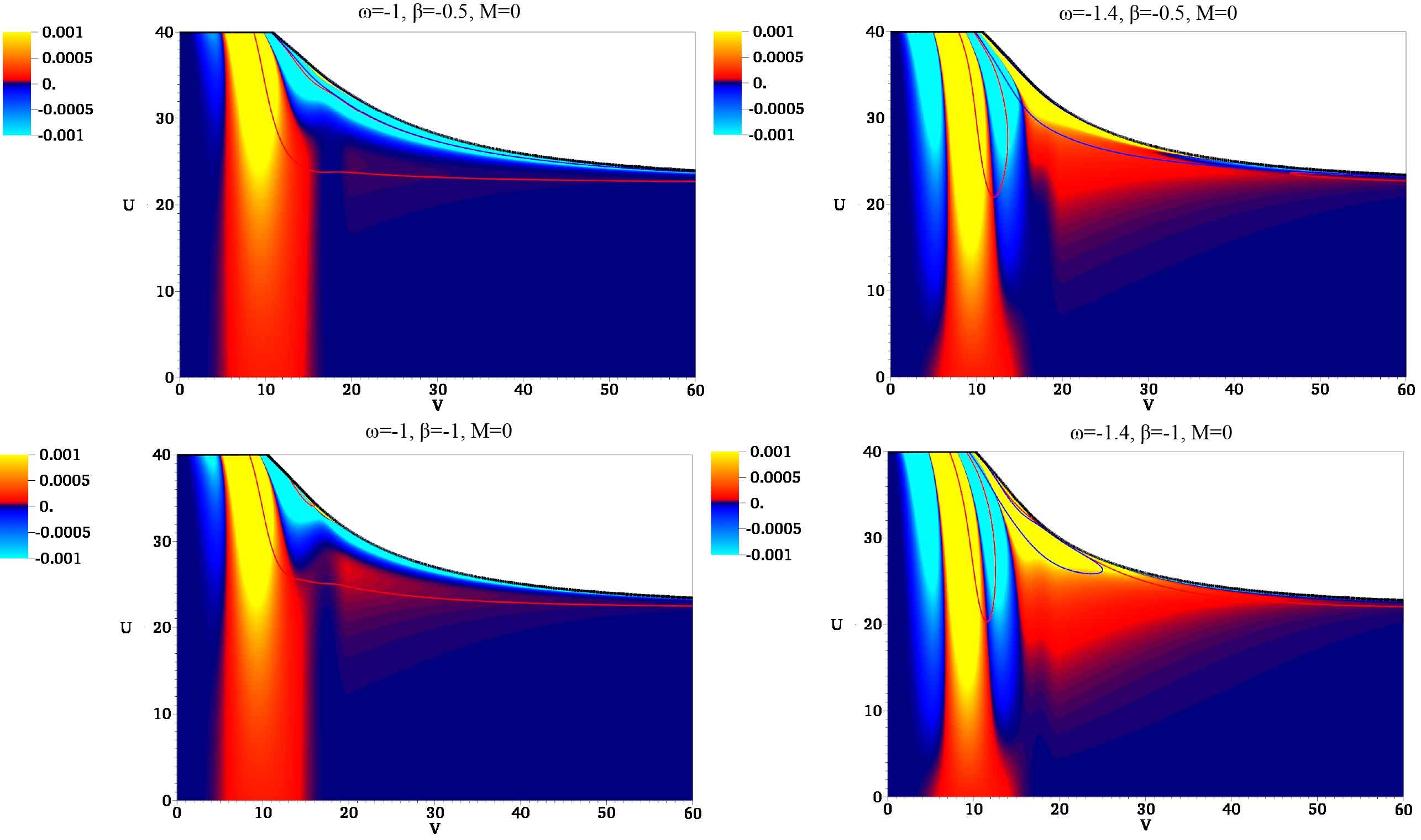}
\caption{\label{fig:Tvv_negativebeta}$T_{vv}$ by varying $\beta = -0.5$ (upper), $-1$ (lower) and varying $\omega = -1$ (left), $-1.4$ (right).}
\end{center}
\end{figure}

\begin{figure}
\begin{center}
\includegraphics[scale=0.14]{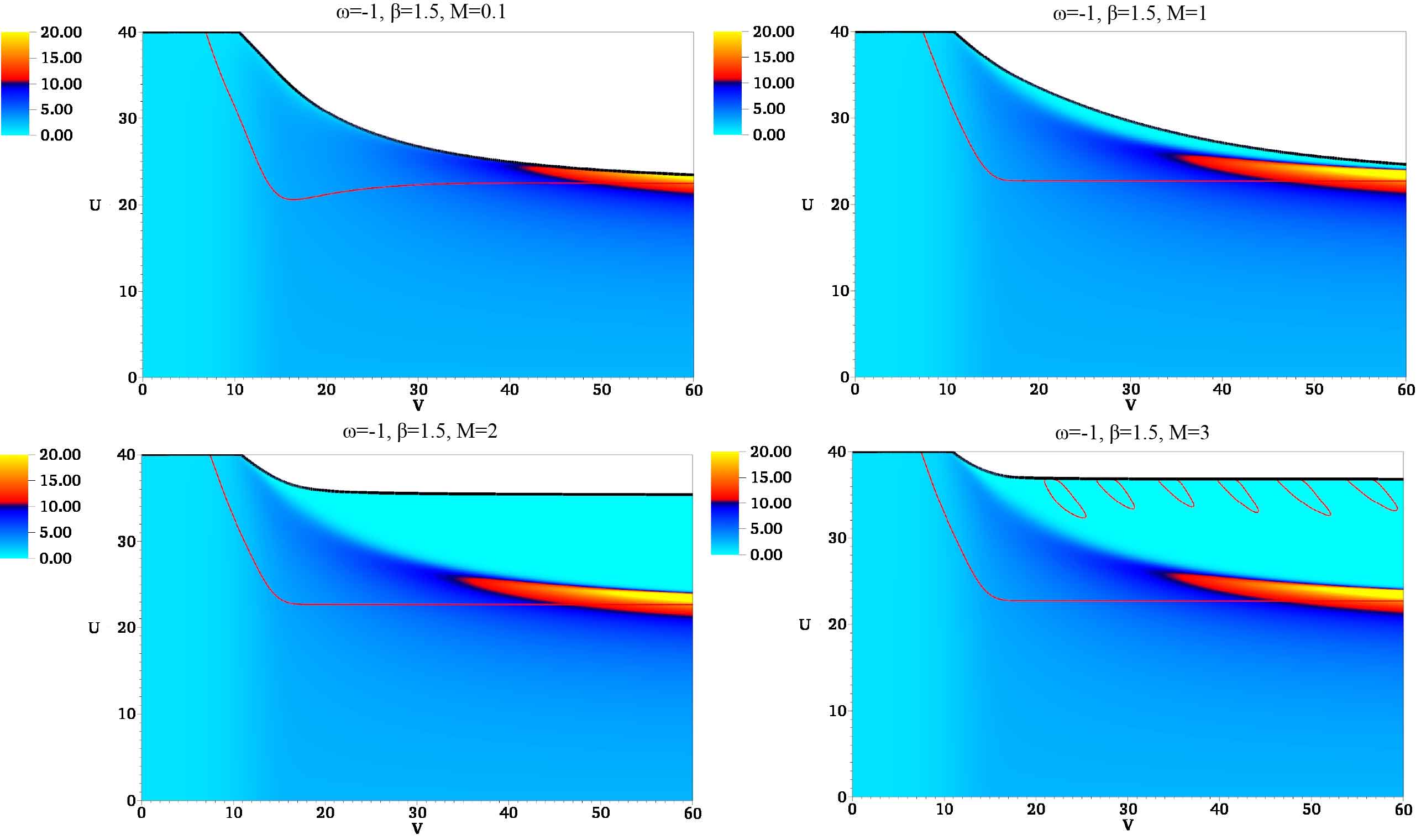}
\caption{\label{fig:alpha_largebeta}$\alpha^{2}$ for $\beta = 1.5$ and varying $M = 0.1, 1, 2,$ and $3$, respectively.}
\end{center}
\end{figure}

\begin{figure}
\begin{center}
\includegraphics[scale=0.14]{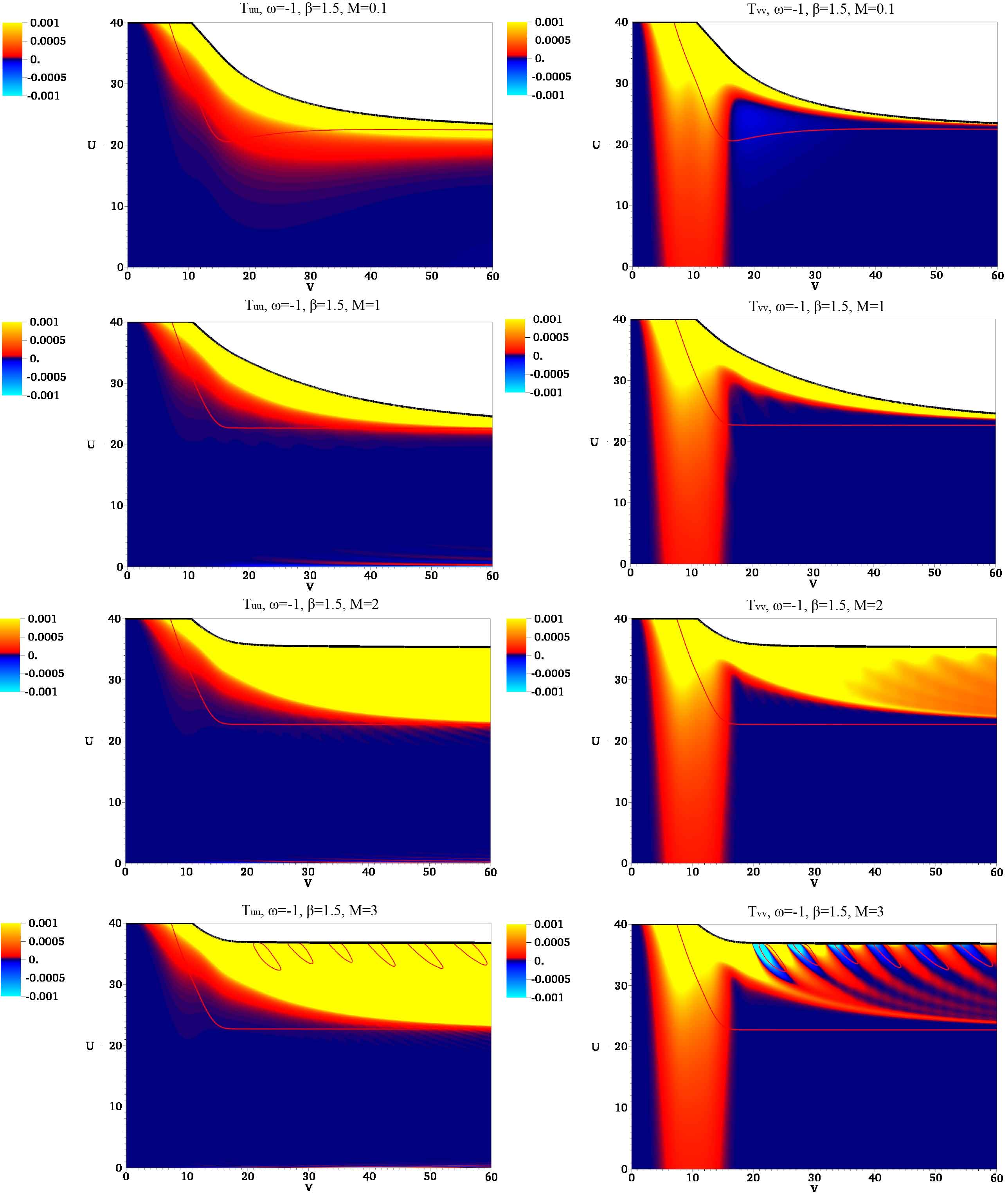}
\caption{\label{fig:T_largebeta}$T_{uu}$ (left) and $T_{vv}$ (right) for $\beta = 1.5$ and varying $M = 0.1, 1, 2,$ and $3$ (from upper to bottom).}
\end{center}
\end{figure}

In this appendix, we summarize numerical data of $\alpha^{2}$, $T_{uu}$, and $T_{vv}$. The conditions that we have used are as follows:
\begin{itemize}
\item[--] String limit, where we use $\beta = 0.5$ and $\beta = 1$, $\omega = -1$ (dilaton limit), and varying the mass parameter $M = 0.1$, $1$, $2$, $3$, and $5$: FIGs.~\ref{fig:alpha_beta}, \ref{fig:Tuu_beta}, and \ref{fig:Tvv_beta}.
\item[--] $\beta < 0$, where we use $\beta = -0.5$ and $-1$, by fixing $M=0$ and varying $\omega = -1$ and $-1.4$: FIGs.~\ref{fig:alpha_negativebeta}, \ref{fig:Tuu_negativebeta}, and \ref{fig:Tvv_negativebeta}.
\item[--] $\beta > 1$, where we use $\beta = 1.5$, by fixing $\omega = -1$ and varying $M = 0.1$, $1$, $2$, and $3$: FIGs.~\ref{fig:alpha_largebeta} and \ref{fig:T_largebeta}.
\end{itemize}
From these data, when there is a Cauchy horizon, we can see that $\alpha$ approaches to zero, where this is an evidence of the existence of mass inflation, as was observed by \cite{Yeom3} (for more justification, see also Appendix A of \cite{Hansen:2014rua}).

\end{document}